\documentclass[aps,prl,twocolumn,preprintnumbers,superscriptaddress,
              showpacs,nofootinbib]{revtex4-1}
\newcommand{\PRE}[1]{}       

\usepackage{bm} \usepackage{epsfig}

\def\beq{\begin{eqnarray}}
\def\eeq{\end{eqnarray}}
\def\bea{\begin{eqnarray}}
\def\eea{\end{eqnarray}}

\newcommand{\gev}{\text{GeV}}

\newcommand{\pb}{\text{pb}}

\newcommand{\eg}{{\em e.g.}}

\newcommand{\eqref}[1]{Eq.~(\ref{#1})}

\newcommand{\figref}[1]{Fig.~\ref{fig:#1}}

\newcommand{\tableref}[1]{Table~\ref{table:#1}}
\newcommand{\tablesref}[2]{Tables~\ref{table:#1} and \ref{table:#2}}

\newcommand{\gsim}{\lower.7ex\hbox{$\;\stackrel{\textstyle>}{\sim}\;$}}
\newcommand{\lsim}{\lower.7ex\hbox{$\;\stackrel{\textstyle<}{\sim}\;$}}

\newcommand{\ssection}[1]{{\bf #1\ }}

\begin{document}

\preprint{UCI-TR-2011-03, UH511-1157-2011}

\title{ \PRE{\vspace*{1.5in}}
Isospin-Violating Dark Matter
\PRE{\vspace*{0.3in}} }

\author{Jonathan L.~Feng}
\affiliation{Department of Physics and Astronomy, University of
California, Irvine, CA 92697, USA
\PRE{\vspace*{.2in}}
}

\author{Jason Kumar}
\affiliation{Department of Physics and Astronomy, University of
Hawaii, Honolulu, HI 96822, USA
\PRE{\vspace*{.2in}}
}

\author{Danny Marfatia}
\affiliation{Department of Physics and Astronomy,
University of Kansas, Lawrence, KS 66045, USA
\PRE{\vspace*{.2in}}
}
\affiliation{Department of Physics, University of Wisconsin, Madison,
WI 53706, USA
\PRE{\vspace*{.4in}}
}

\author{David Sanford\PRE{\vspace*{.3in}}}
\affiliation{Department of Physics and Astronomy, University of
California, Irvine, CA 92697, USA
\PRE{\vspace*{.2in}}
}

\begin{abstract}
\PRE{\vspace*{.3in}} Searches for dark matter scattering off nuclei
are typically compared assuming that the dark matter's
spin-independent couplings are identical for protons and neutrons.
This assumption is neither innocuous nor well motivated.  We consider
isospin-violating dark matter (IVDM) with one extra parameter, the
ratio of neutron to proton couplings, and include the isotope
distribution for each detector. For a single choice of the coupling
ratio, the DAMA and CoGeNT signals are consistent with each other and
with current XENON constraints, and they unambiguously predict near
future signals at XENON and CRESST.  We provide a quark-level
realization of IVDM as WIMPless dark matter that is consistent with
all collider and low-energy bounds.
\end{abstract}


\maketitle

\ssection{Introduction.} Dark matter makes up five-sixths of the
matter in the Universe, but all current evidence for dark matter is
through its gravitational effects.  The detection of dark matter
scattering through non-gravitational interactions would be a large
step toward identifying dark matter, and there are many experiments
searching for such events.  The excitement around this approach has
been heightened recently by data from the DAMA~\cite{Bernabei:2008yi}
and CoGeNT~\cite{Aalseth:2010vx} experiments, which are consistent
with scattering by a dark matter particle with mass $m_X \sim 10~\gev$
and spin-independent (SI) $X$-nucleon scattering cross sections
$\sigma_N \sim 2\times 10^{-4}~\pb$ and $5\times 10^{-5}~\pb$,
respectively.  This excitement is, however, tempered by null results
from XENON~\cite{Aprile:2011hi,Angle:2011th} and
CDMS~\cite{Akerib:2010pv,Ahmed:2010wy}, leaving a confusing picture
that has motivated much theoretical and experimental work.

The comparison of dark matter experimental results is subject to an
array of assumptions and uncertainties from particle physics, nuclear
physics, and astrophysics.  In this study, we focus on a particularly
simple and common particle physics assumption, that of flavor isospin
invariance.  Dark matter detectors have various nuclear compositions.
To derive implications for $\sigma_N$, experiments almost universally
assume that dark matter couples identically to protons and neutrons.
This assumption is not well motivated. For example, Dirac neutrinos
and sneutrinos have isospin-violating couplings, and, in fact, even
neutralino couplings are generically isospin-violating, although
typically at insignificant levels (see, \eg,
Refs.~\cite{Ellis:2001hv,Cotta:2009zu}).  In any case, as these and
other conventional dark matter candidates do not easily explain the
DAMA and CoGeNT signals, it is reasonable to consider more general
frameworks.  Here we consider flavor isospin-violating dark matter
(IVDM) with one extra parameter, the ratio of neutron to proton
couplings $f_n/f_p$.

IVDM has been considered previously in general
analyses~\cite{Kurylov:2003ra}, and also recently in studies of
various interpretations of the CoGeNT
results~\cite{Chang:2010yk,Kang:2010mh}. We focus solely on IVDM and
consider for the first time the distribution of isotopes present in
each detector.  Previous work has neglected this distribution, which
implies that dark matter may be completely decoupled from any given
detector for a particular value of $f_n/f_p$.  However, this is not
true if there is more than one isotope present, as is the case in many
detectors, and the viability and implications of IVDM cannot be
established without considering the isotope distribution.  As we will
see, including the isotope distribution has remarkable consequences.
For a single choice of $f_n/f_p$, the DAMA and CoGeNT signals are
consistent with each other and with current XENON constraints.  At the
same time, the isotope distribution implies that XENON cannot be
completely decoupled, and the IVDM scenario unambiguously predicts
near future signals at XENON and other detectors, such as CRESST and
COUPP. We identify and discuss slight inconsistencies with other data,
and present a general analysis of when experiments may be reconciled
by isospin violation. Finally, we provide a quark-level realization of
IVDM as WIMPless dark matter~\cite{Feng:2008ya,Feng:2008dz} that is
consistent with all collider and low-energy bounds.

\ssection{Cross sections for IVDM.} We focus on the SI scattering of
an IVDM particle $X$ off a nucleus $A$ with $Z$ protons and $A-Z$
neutrons.  The event rate is
\begin{equation}
R =N_T n_X  \! \! \int \!  d E_R \!
\int_{v_{\text{min}}}^{v_{\text{max}}} \! \! \!  d^3 v \, f(v) v
\frac{d\sigma}{dE_R} \ ,
\end{equation}
where $N_T$ is the number of target nuclei, $n_X$ is the local number
density of dark matter particles, and the limits of the recoil energy
$E_R$ integral are determined by experimental considerations. The IVDM
particle's velocity $v$ varies from $v_{\text{min}} = \sqrt{ m_A E_R /
2 \mu_A^2 }$, where $\mu_A = m_A m_X / (m_A + m_X)$, to
$v_{\text{max}}$, a function of the halo escape velocity, and $f(v)$
is the distribution of $X$ velocities relative to the detector.  The
differential cross section is $d\sigma/dE_R = \hat \sigma_A m_A / (2
v^2 \mu_A^2)$, with
\begin{equation}
\hat \sigma_A \! = \! \frac{\mu_A^2}{M_*^4}
\left[ f_p Z F_A^p (E_R) + f_n (A \! - \! Z) F_A^n (E_R) \right]^2 ,
\end{equation}
where $f_{p,n}$ are the couplings to protons and neutrons, normalized
by the choice of mass scale $M_*$, and $F_A^{p,n} (E_R)$ are the
proton and neutron form factors for nucleus $A$.

$F_A^p (E_R)$ and $F_A^n (E_R)$ are not identical. $F_A^p (E_R)$ is
what has typically been measured, but $F_A^n (E_R)$ may also be
probed, for example, through neutrino and electron parity-violating
scattering off nuclei~\cite{Amanik:2009zz}.  However, since the
isospin violation from this effect is small compared to the
potentially large effects of varying $f_n/f_p$, we will set both form
factors equal to $F_A (E_R)$.  With this approximation, the event rate
simplifies to $R = \sigma_A I_A$, where
\begin{eqnarray}
\label{SIcross}
\sigma_A &=& \frac{\mu_A^2}{M_*^4}
\left[ f_p Z + f_n (A-Z) \right]^2 \\
I_A &=&N_T n_X\! \! \int \!  d E_R \!
\int_{v_{\text{min}}}^{v_{\text{max}}} \! \! \!  d^3 v \,
f(v)  \frac{m_A}{2 v \mu_A^2} F_A^2 (E_R) \, ,
\end{eqnarray}
and $\sigma_A$ is the zero-momentum-transfer SI cross section from
particle physics, and $I_A$ depends on experimental, astrophysical,
and nuclear physics inputs.  If $f_n = f_p$, we recover the well-known
relation $R \propto A^2$.  For IVDM, however, the scattering
amplitudes for protons and neutrons may interfere destructively, with
complete destructive interference for $f_n/f_p = -Z/(A-Z)$.

We assume that each detector either has only one element, or that the
recoil spectrum allows one to distinguish one element as the dominant
scatterer.  But it is crucial to include the possibility of multiple
isotopes.  The event rate is then $R = \sum_i \eta_i \sigma_{A_i}
I_{A_i}$, where the sum is over isotopes $A_i$ with fractional number
abundance $\eta_i$.

\ssection{IVDM and current data.} It will be convenient to define two
nucleon cross sections.  The first is $\sigma_p = \mu_p^2 f_p^2 /
M_*^4$, {\em the $X$-proton cross section}.  In terms of $\sigma_p$,
\begin{equation}
R = \sigma_p \sum_i \eta_i \frac{\mu_{A_i}^2}{\mu_p^2}
I_{A_i} \left[Z + (A_i-Z) f_n / f_p \right]^2 .
\end{equation}
The second is $\sigma_N^Z$, {\em the typically-derived $X$-nucleon
cross section from scattering off nuclei with atomic number $Z$,
assuming isospin conservation and the isotope abundances found in
nature}.  With the simplification that the $I_{A_i}$ vary only mildly
for different $i$, we find
\begin{equation}
\frac{\sigma_p}{\sigma_N^Z}
= \frac{\sum_i \eta_i \mu_{A_i}^2 A_i^2}
{\sum_i \eta_i \mu_{A_i}^2 [Z + (A_i - Z) f_n/f_p]^2}
\equiv F_Z  \ .
\end{equation}
If one isotope dominates, the well-known result, $F_Z = [Z/A + (1-Z/A)
f_n/f_p]^{-2}$, is obtained.

\begin{figure}[tb]
\includegraphics[width=0.95\columnwidth]{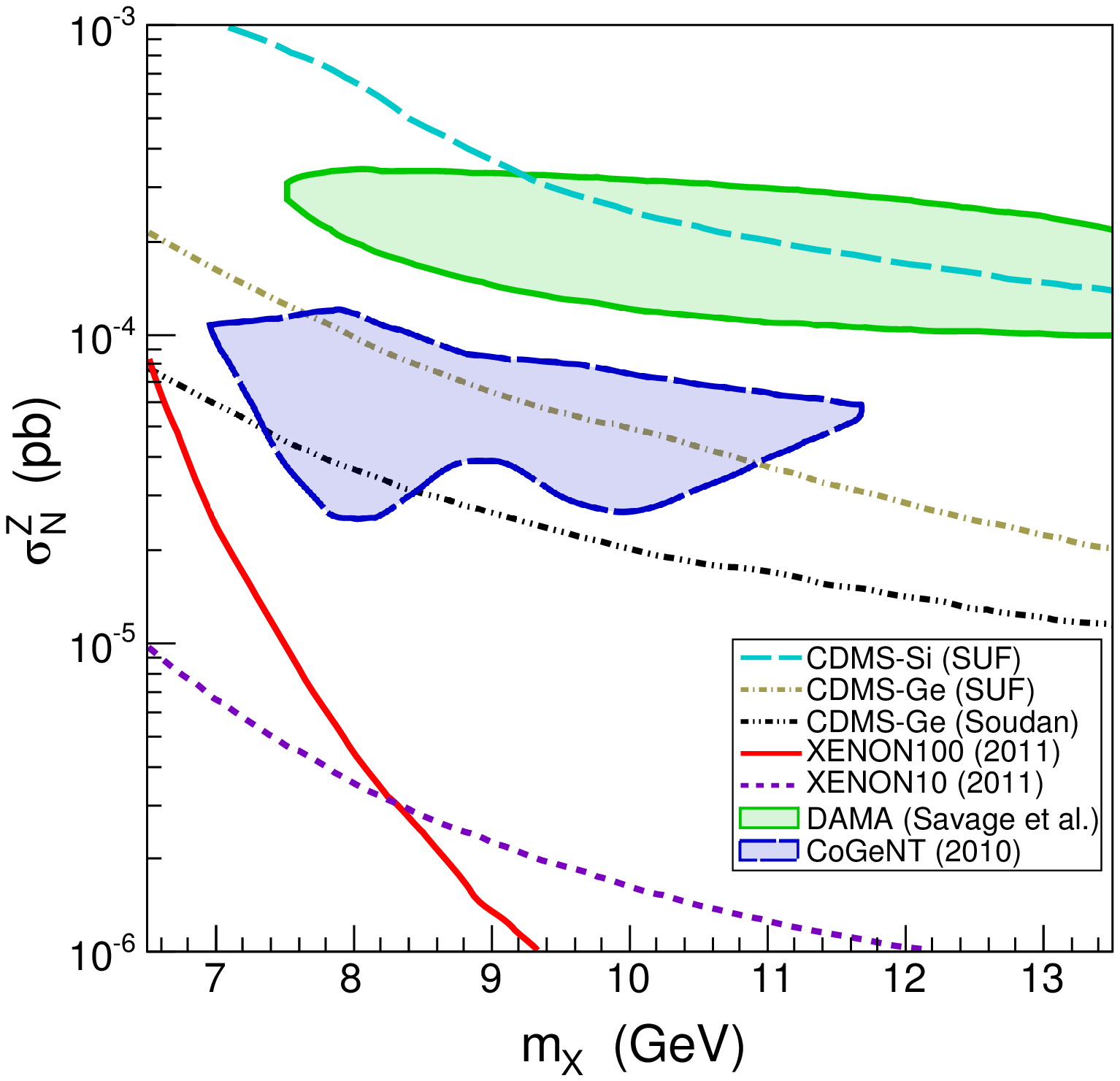}
\includegraphics[width=0.95\columnwidth]{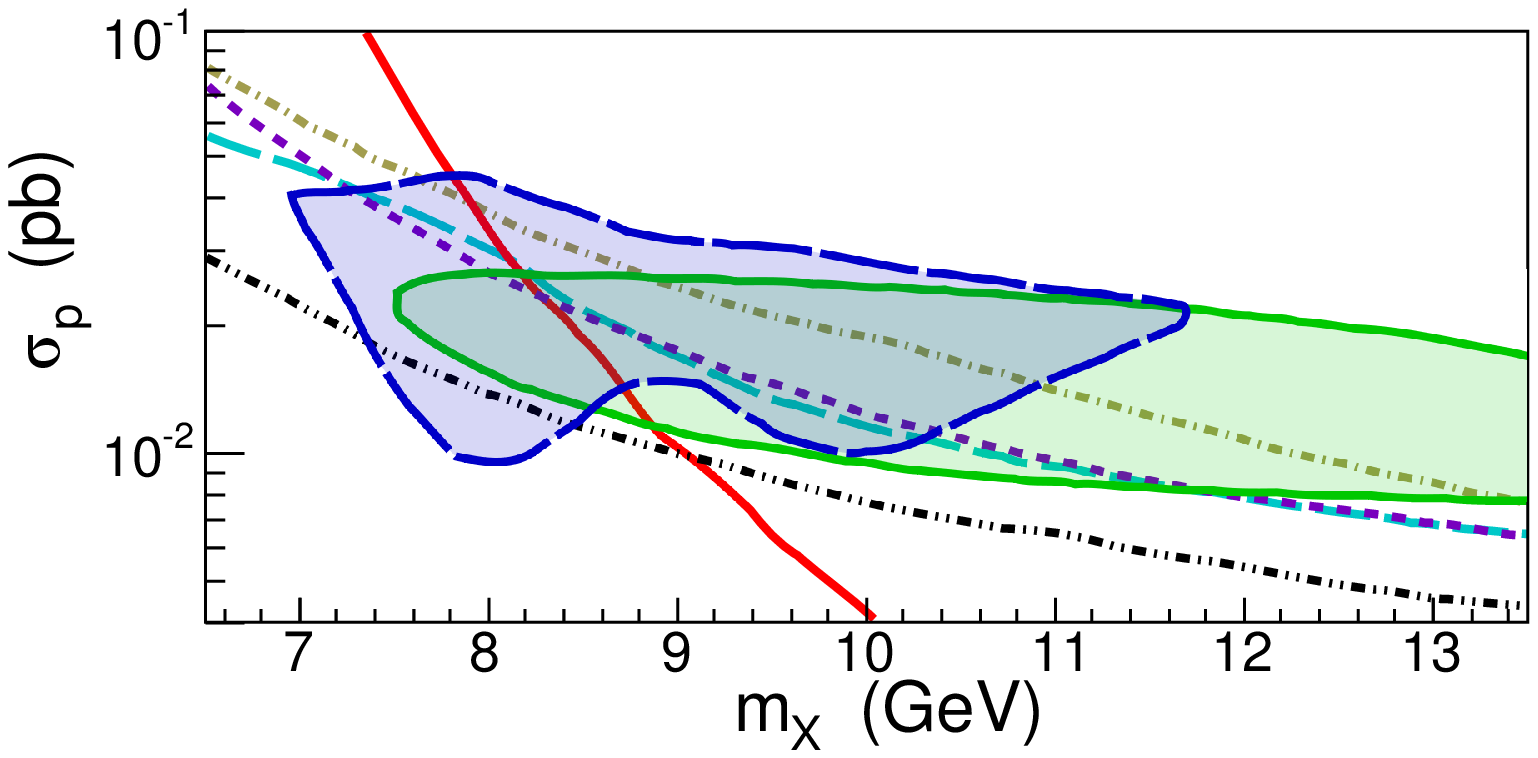}
\vspace*{-.1in}
\caption{\label{fig:DDbounds} Favored regions and exclusion contours
in the $(m_X, \sigma_N^Z)$ plane (top), and in the $(m_X, \sigma_p)$
plane for IVDM with $f_n / f_p = - 0.7$ (bottom).  }
\end{figure}

In \figref{DDbounds} we show regions in the $(m_X, \sigma_N^Z)$ plane
and the $(m_X, \sigma_p)$ plane for $f_n/f_p = -0.7$ that are favored
and excluded by current bounds.  These include the DAMA 3$\sigma$
favored region~\cite{Savage:2008er,Savage:2010tg}, assuming no
channeling~\cite{Bozorgnia:2010xy} and that the signal arises entirely
from Na scattering; the CoGeNT 90\% CL favored
region~\cite{Aalseth:2010vx}; 90\% CL exclusion contours from
XENON100~\cite{Aprile:2011hi} and XENON10~\cite{Angle:2011th}; and
90\% CL bounds from CDMS Ge and Si~\cite{Akerib:2010pv,Ahmed:2010wy}.
The isotope abundances are given in \tablesref{tableofR}{tableofA}.

There are controversies regarding the exclusion contours for
xenon-based detectors at low mass~\cite{leffandpoisson}.  The energy
dependence of the scintillation efficiency at low energies is
uncertain, and there are questions about the assumption of Poisson
fluctuations in the expected photoelectron count for light dark
matter.  We have also not accounted for uncertainties in the
associated quenching factors for Na, Ge and Si~\cite{Hooper:2010uy}.
These issues can enlarge some of the signal regions or alter some of
the exclusion curves of \figref{DDbounds}.  We have also not adjusted
the favored regions and bounds to account for differences in the dark
matter velocity distributions adopted by the various analyses, which
would slightly shift the contours.

\begin{table}[tb]
\caption{$R_{\text{max}}[Z_1 , Z_2]$, where the $Z_1$ ($Z_2$) elements
are listed in rows (columns).  Elements with one significant isotope
have their $(Z, A)$ listed; those with more than one are denoted by
asterisks and listed in \tableref{tableofA}.}
\small{ \begin{tabular}{c|rrrrrrr}
Element & Xe \quad & Ge \quad  & Si  \quad & Ca \quad 
& W  \quad & Ne  \quad & C \qquad \\
\hline
Xe\,(54,*) & 1.00 & 8.79 & 149.55 & 138.21 & 10.91 & 34.31 & 387.66 \\
Ge\,(32,*) & 22.43 & 1.00 & 68.35 & 63.14 & 130.45 & 15.53 & 176.47 \\
Si\,(14,*) & 172.27 & 30.77 & 1.00 & 1.06 & 757.44 & 1.06 & 2.67 \\
Ca\,(20,*) & 173.60 & 31.53 & 1.17 & 1.00 & 782.49 & 1.10 & 2.81 \\
W\,(74,*) & 2.98 & 13.88 & 177.46 & 166.15 & 1.00 & 41.64 & 466.75 \\
Ne\,(10,*) & 163.65 & 28.91 & 4.39 & 4.09 & 726.09 & 1.00 & 11.52 \\
C\,(6,*) & 176.35 & 32.13 & 1.07 & 1.02 & 789.59 & 1.12 & 1.00 \\
I\,(53,127) & 1.94 & 5.51 & 127.04 & 118.35 & 20.68 & 28.92 & 326.95 \\
Cs\,(55,133) & 1.16 & 7.15 & 139.65 & 127.61 & 12.32 & 31.88 & 355.27 \\
O\,(8,16) & 178.49 & 32.13 & 1.08 & 1.03 & 789.90 & 1.13 & 1.01 \\
Na\,(11,23) & 101.68 & 13.77 & 8.45 & 8.33 & 481.03 & 2.27 & 22.68 \\
Ar\,(18,40) & 35.38 & 1.87 & 45.39 & 42.56 & 190.26 & 10.32 & 119.80 \\
F\,(9,19) & 89.39 & 10.88 & 12.44 & 11.90 & 425.93 & 3.05 & 33.47
\end{tabular}
}
\label{table:tableofR}
\end{table}

\begin{table}[tb]
\caption{$A_i$ for isotopes and their fractional number abundances
$\eta_i$ in percent for all isotopes with $\eta_i > 1\%$.}
\begin{tabular}{lllllll}
\quad Xe & \quad Ge & \quad Si & \quad Ca & \quad W & \quad Ne & \quad
C \\
\hline
128\,(1.9) & 70\,(21) & 28\,(92) & 40\,(97) & 182\,(27) & 20\,(91) 
  & 12\,(99) \\
129\,(26) & 72\,(28) & 29\,(4.7) & 44\,(2.1) & 183\,(14) & 22\,(9.3) 
  & 13\,(1.1) \\
130\,(4.1) & 73\,(7.7) & 30\,(3.1) & & 184\,(31) \\
131\,(21) & 74\,(36) & & & 186\,(28) \\
132\,(27) & 76\,(7.4) \\
134\,(10)\\
136\,(8.9)
\end{tabular}
\label{table:tableofA}
\end{table}

Remarkably, for $-0.72 \alt f_n/f_p \alt -0.66$, the DAMA- and
CoGeNT-favored regions overlap {\em and} the sensitivity of XENON is
sufficiently reduced to be consistent with these signals, since this
choice of $f_n/f_p$ leads to nearly complete destructive interference
for the proton/neutron content of xenon isotopes. The possibility of
IVDM therefore brings much of the world's data into agreement and
leads to a very different picture than that implied by studies
assuming isospin conservation.  The CDMS Ge constraint marginally
excludes the overlapping region, and since CoGeNT utilizes Ge, the
tension between CoGeNT and CDMS Ge cannot be alleviated by isospin
violation.  However, it is possible that an improved understanding of
CoGeNT backgrounds and the energy scale calibration of the CDMS Ge
detectors at low energy may resolve the
disagreement~\cite{Hooper:2010uy,Collar:2011kf,Ahmed:2010wy}.

\ssection{Predictions.} Further tests of the IVDM hypothesis may come
from other detectors.  If two experiments report signals suggesting
the same $m_X$, their results imply an experimental measurement of
\begin{equation}
R [Z_1 , Z_2] \equiv \sigma_N^{Z_1} / \sigma_N^{Z_2} \ .
\end{equation}
$R [Z_1, Z_2] = F_{Z_2}/F_{Z_1}$ is then a quadratic equation in
$f_n/f_p$, the solution of which enables unambiguous signal
predictions for other detectors.

As timely examples, consider two current experiments.  Preliminary
results from CRESST may indicate a signal from scattering off
oxygen~\cite{CRESST}.  $f_n /f_p \approx -0.7$ implies $F_{Z=8}
\approx 44$.  The IVDM explanation of DAMA and CoGeNT therefore
predicts that CRESST will see a signal consistent with $m_X \sim
10~\gev$ and $\sigma_N^{Z=8} \sim 8.5\, \sigma_N^{Z=32}$. Such a cross
section may in fact be consistent with CRESST
data~\cite{Hooper:2010uy,CRESSTII}.  COUPP is a ${\rm CF_3 I}$
detector; its sensitivity to low-mass dark matter arises from C and F
scattering.  For $f_n /f_p \approx -0.7$, $m_X \sim 10~\gev$, we find
$\sigma_N^{Z=6} \sim 8.4\, \sigma_N^{Z=32}$ and $\sigma_N^{Z=9} \sim
4.2\, \sigma_N^{Z=32}$.  COUPP would be expected to report a
normalized cross section between these values, with the value
depending on the relative detection power of the C and F targets.

\ssection{Relative detection prospects.} Although XENON excludes
CoGeNT and DAMA signals assuming isospin conservation, this is not the
case for IVDM.  One might then ask: given any signal at a detector
with atomic number $Z_1$, what sensitivity is required for a detector
with atomic number $Z_2$ to either corroborate or disfavor this
signal, allowing for isospin violation?  Maximizing $R[Z_1 , Z_2]$
with respect to $f_n / f_p$ determines the factor by which the $Z_2$
detector must exclude the $Z_1$ signal assuming isospin conservation,
such that the $Z_1$ signal is excluded even allowing for isospin
violation.  Similarly, maximizing $R[Z_2, Z_1]$ determines the factor
by which the $Z_2$ detector may come up short in probing an
isospin-conserving origin for the $Z_1$ signal, while still having the
potential to find evidence for an isospin-violating origin.

In \tableref{tableofR}, we present $R_{\text{max}}[Z_1, Z_2]$, the
maximal value of $R[Z_1 , Z_2]$ for all possible values of $f_n/f_p$,
for many materials that are commonly used in dark matter detectors.
The isotope composition of elements plays an important role in
determining $R_{\text{max}}[Z_1, Z_2]$.  If the element $Z_2$ is
composed entirely of one isotope, then it is always possible to choose
$f_n / f_p$ so that $\sigma_{Z_2} = 0$ and thus $R_{\text{max}}[Z_1
\neq Z_2, Z_2 ] = \infty $; these columns have been omitted from
\tableref{tableofR}.  However, if there is more than one significant
isotope, it is impossible to achieve exact destructive interference
for all isotopes simultaneously, and so $R_{\text{max}}[Z_1 \neq Z_2,
Z_2 ]$ is finite.  In particular, although isospin violation can
weaken the bounds achieved by Xe and Ge detectors, we see in
\tableref{tableofR} that these bounds can be weakened by at most two
orders of magnitude.  Upcoming XENON results may therefore exclude
DAMA and CoGeNT, even for IVDM; XENON bounds already eliminate some of
the DAMA/CoGeNT overlap region (\figref{DDbounds}), and will probe the
entire region if XENON sensitivities are improved by an order of
magnitude.

\ssection{Isospin violation in WIMPless models.} So far we have worked
at the nucleon level.  We now provide a quark-level theory of dark
matter that generically realizes isospin violation.  In supersymmetric
WIMPless dark matter models~\cite{Feng:2008ya,Feng:2008dz}, dark
matter particles $X$ freeze out in a hidden sector with the correct
relic density and interact with the standard model through connector
particles $Y$.  We consider the superpotential
\begin{equation}
W =\sum_i (\lambda_q^i X Y_{q_L} q_L^i + \lambda_u^i X Y_{u_R} u_R^i
+ \lambda_d^i X Y_{d_R} d_R^i ) \,,
\end{equation}
where $X$ is a real scalar dark matter particle, $q_L, u_R, d_R$ are
standard model quarks, $i$ labels generations, and the connectors
$Y_{q_L,u_R,d_R}$ are 4$^{\text{th}}$ generation mirror quarks.
Assuming real Yukawa couplings and $m_Y = m_{Y_{u,d}} \gg m_X ,m_q$,
the connector particles induce the SI operators
\begin{equation}
\label{opeq}
{\cal O}_i = \lambda_q^i \lambda_u^i X X \bar u^i u^i / m_Y
+ \lambda_q^i \lambda_d^i X X \bar d^i d^i / m_Y\,,
\end{equation}
leading to the scattering cross section of \eqref{SIcross} with $
f_{p,n} / M_*^2 = \sum_i (\lambda_q^i \lambda_u^i B_{u^i}^{p,n}
+\lambda_q^i \lambda_d^i B_{d^i}^{p,n}) / (\sqrt{\pi} m_X m_Y)$.  The
$B_{q^i}^{p,n}$ are integrated nuclear form factors, including $B_u^p
= B_d^n \approx 6$, $B_u^n = B_d^p \approx 4$~\cite{Ellis:2001hv}.

The amount of isospin violation in dark matter-nucleus interactions is
solely determined by the Yukawa flavor structure.  There are many
possibilities; WIMPless models may explain the DAMA signal with
couplings to either the 1$^{\text{st}}$~\cite{Feng:2008ya} or
3$^{\text{rd}}$~\cite{Feng:2008dz,Zhu:2011dz} generation.  Here we
assume only 1$^{\text{st}}$ generation quark couplings, automatically
satisfying flavor constraints.  Assuming $m_X = 10~\gev$ and $m_Y =
400~\gev$, consistent with all collider and precision electroweak
bounds, the region of the $(\lambda_q^1 \lambda_u^1, \lambda_q^1
\lambda_d^1)$ plane that explains DAMA and CoGeNT is
\begin{eqnarray}
\label{regioneq}
\lambda_u^1 \simeq -1.08 \, \lambda_d^1,  \qquad
0.013 \lsim \lambda_q^1 \lambda_d^1 \lsim 0.024 \ .
\end{eqnarray}
IVDM is clearly generic in this microscopic model of dark matter
interactions and may simultaneously reconcile the DAMA and CoGeNT
signals and XENON bounds.

The IVDM reconciliation of DAMA, CoGeNT, and XENON relies on
cancellations between $p$ and $n$ couplings, and so requires larger
couplings than in the isospin-preserving case to maintain the desired
DAMA and CoGeNT signals.  Such models may potentially violate collider
constraints, which are not subject to cancellations.  This WIMPless
model provides a quark-level framework in which one may investigate
this question.

The most stringent model-independent constraints are from Tevatron
searches for $p \bar p \to XX +
\text{jet}$~\cite{Birkedal:2004xn,Goodman:2010ku}.  Using
MadGraph/MadEvent 4.4.32~\cite{Alwall:2007st}, one can compute the
monojet cross section (requiring jet $E_T > 80~\gev$) induced by the
operator of \eqref{opeq}.  The resulting 2$\sigma$ bounds from
Tevatron data are roughly $\lambda_q^1 \lambda_{u,d}^1 \alt 1$, two
orders of magnitude too weak to probe the DAMA and CoGeNT favored
couplings described in \eqref{regioneq}.

\ssection{Conclusions.} Results for spin-independent dark matter
interactions typically assume identical couplings to protons and
neutrons.  Isospin violation is generic, however, and we have shown
that IVDM with $f_n/f_p \approx -0.7$ may explain both DAMA and
CoGeNT, consistent with XENON10/100 bounds.  This scenario is only
marginally excluded by CDMS Ge constraints, unambiguously predicts a
signal at CRESST, and may even be tested by XENON, given its several
significant isotopes, as discussed above; near future data will shed
light on this picture.  More generally, we have explored the extent to
which dropping the $f_p=f_n$ assumption may reconcile results from
various detectors, stressing the important role played by the
distribution of isotopes. Finally, we have shown that IVDM is easily
realized in a quark-level model consistent with all low-energy and
collider observables.

\ssection{Acknowledgments.} We gratefully acknowledge discussions with
A.~Rajaraman, P.~Sandick, W.~Shepherd, P.~Sorensen, S.~Su, T.~Tait,
and X.~Tata.  JLF and DS are supported in part by NSF grants
PHY-0653656 and PHY-0970173.  JK is supported by DOE grant
DE-FG02-04ER41291.  DM is supported in part by DOE grant
DE-FG02-04ER41308 and NSF Grant PHY-0544278.

\ssection{Note added.} After the completion of this work, an annual
modulation signal from CoGeNT and a new constraint from SIMPLE have
been reported.  These results and some of the following discussion may
be found in Refs.~\cite{Aalseth:2011wp,Felizardo:2011uw}.


\begin{thebibliography}{99}

\bibitem{Bernabei:2008yi}
  R.~Bernabei {\it et al.}  [DAMA Collaboration],
  Eur.\ Phys.\ J.\  C {\bf 56}, 333 (2008)
  [arXiv:0804.2741 [astro-ph]].

\bibitem{Aalseth:2010vx}
  C.~E.~Aalseth {\it et al.}  [CoGeNT Collaboration],
  arXiv:1002.4703 [astro-ph.CO].

\bibitem{Aprile:2011hi}
  E.~Aprile {\it et al.}  [XENON100 Collaboration],
  arXiv:1104.2549 [astro-ph.CO].

\bibitem{Angle:2011th}
  J.~Angle {\it et al.}  [XENON10 Collaboration],
  arXiv:1104.3088 [astro-ph.CO].

\bibitem{Akerib:2010pv}
  D.~S.~Akerib {\it et al.}  [CDMS Collaboration],
  Phys.\ Rev.\  D {\bf 82}, 122004 (2010)
  [arXiv:1010.4290 [astro-ph.CO]].

\bibitem{Ahmed:2010wy}
  Z.~Ahmed {\it et al.}  [CDMS-II Collaboration],
  Phys.\ Rev.\ Lett.{\bf 106}, 131302 (2011)
  [arXiv:1011.2482v3 [astro-ph.CO]].

\bibitem{Ellis:2001hv}
  J.~R.~Ellis, J.~L.~Feng, A.~Ferstl, K.~T.~Matchev and K.~A.~Olive,
  Eur.\ Phys.\ J.\  C {\bf 24}, 311 (2002)
  [arXiv:astro-ph/0110225].

\bibitem{Cotta:2009zu}
  R.~C.~Cotta, J.~S.~Gainer, J.~L.~Hewett and T.~G.~Rizzo,
  New J.\ Phys.\  {\bf 11}, 105026 (2009)
  [arXiv:0903.4409 [hep-ph]].

\bibitem{Kurylov:2003ra}
  A.~Kurylov and M.~Kamionkowski,
  Phys.\ Rev.\  D {\bf 69}, 063503 (2004)
  [arXiv:hep-ph/0307185];
  F.~Giuliani,
  Phys.\ Rev.\ Lett.\  {\bf 95}, 101301 (2005)
  [arXiv:hep-ph/0504157].

\bibitem{Chang:2010yk}
  S.~Chang, J.~Liu, A.~Pierce, N.~Weiner and I.~Yavin,
  JCAP {\bf 1008}, 018 (2010)
  [arXiv:1004.0697 [hep-ph]].

\bibitem{Kang:2010mh}
  Z.~Kang, T.~Li, T.~Liu, C.~Tong and J.~M.~Yang,
  JCAP {\bf 1101}, 028 (2011)
  [arXiv:1008.5243 [hep-ph]].

\bibitem{Feng:2008ya}
J.~L.~Feng and J.~Kumar,
  Phys.\ Rev.\ Lett. {\bf 101}, 231301 (2008)
  [arXiv:0803.4196 [hep-ph]];
J.~L.~Feng, H.~Tu, and H.~B.~Yu,
 JCAP {\bf 0810}, 043 (2008)
  [arXiv:0808.2318 [hep-ph]].

\bibitem{Feng:2008dz}
  J.~L.~Feng, J.~Kumar and L.~E.~Strigari,
  Phys.\ Lett.\  B {\bf 670}, 37 (2008)
  [arXiv:0806.3746 [hep-ph]].

\bibitem{Amanik:2009zz}
  P.~S.~Amanik and G.~C.~McLaughlin,
  J.\ Phys.\ G {\bf 36}, 015105 (2009);
  S.~Ban, C.~J.~Horowitz and R.~Michaels,
  arXiv:1010.3246 [nucl-th].

\bibitem{Savage:2008er}
  C.~Savage, G.~Gelmini, P.~Gondolo and K.~Freese,
  JCAP {\bf 0904}, 010 (2009)
  [arXiv:0808.3607 [astro-ph]].

\bibitem{Savage:2010tg}
  C.~Savage, G.~Gelmini, P.~Gondolo and K.~Freese,
  Phys.\ Rev.\  {\bf D83}, 055002 (2011)
  [arXiv:1006.0972 [astro-ph.CO]].

\bibitem{Bozorgnia:2010xy}
  N.~Bozorgnia, G.~B.~Gelmini and P.~Gondolo,
  JCAP {\bf 1011}, 019 (2010)
 [arXiv:1006.3110 [astro-ph.CO]].

\bibitem{leffandpoisson}
  J.~I.~Collar and D.~N.~McKinsey,
  [arXiv:1005.0838 [astro-ph.CO]];
  XENON100 Collaboration,
  [arXiv:1005.2615 [astro-ph.CO]];
  J.~I.~Collar and D.~N.~McKinsey,
  [arXiv:1005.3723 [astro-ph.CO]].

\bibitem{Hooper:2010uy}
  D.~Hooper, J.~I.~Collar, J.~Hall, D.~McKinsey and C.~Kelso,
  Phys.\ Rev.\  D {\bf 82}, 123509 (2010)
  [arXiv:1007.1005 [hep-ph]].

\bibitem{Collar:2011kf}
  J.~I.~Collar,
  arXiv:1103.3481 [astro-ph.CO].

\bibitem{CRESST}
See talk by W.~Seidel at IDM2010.

\bibitem{CRESSTII}
See talk by T.~Schwetz at IDM2010.

\bibitem{Zhu:2011dz}
  G.~Zhu,
  Phys.\ Rev.\  {\bf D83}, 076011 (2011).
  [arXiv:1101.4387 [hep-ph]].

\bibitem{Birkedal:2004xn}
  A.~Birkedal, K.~Matchev and M.~Perelstein,
  Phys.\ Rev.\  D {\bf 70}, 077701 (2004)
  [arXiv:hep-ph/0403004];
  J.~L.~Feng, S.~Su and F.~Takayama,
  Phys.\ Rev.\ Lett.\  {\bf 96}, 151802 (2006)
  [arXiv:hep-ph/0503117].

\bibitem{Goodman:2010ku}
  J.~Goodman, M.~Ibe, A.~Rajaraman, W.~Shepherd, T.~M.~P.~Tait and H.~B.~Yu,
  Phys.\ Rev.\  {\bf D82}, 116010 (2010)
  [arXiv:1008.1783 [hep-ph]].

\bibitem{Alwall:2007st}
  J.~Alwall {\it et al.},
  JHEP {\bf 0709}, 028 (2007)
  [arXiv:0706.2334 [hep-ph]].

\bibitem{Aalseth:2011wp}
  C.~E.~Aalseth {\it et al.},
  arXiv:1106.0650 [astro-ph.CO];
  T.~Schwetz and J.~Zupan,
  arXiv:1106.6241 [hep-ph];
  M.~Farina, D.~Pappadopulo, A.~Strumia and T.~Volansky,
  arXiv:1107.0715 [hep-ph];
  P.~J.~Fox, J.~Kopp, M.~Lisanti and N.~Weiner,
  arXiv:1107.0717 [hep-ph].

\bibitem{Felizardo:2011uw}
  M.~Felizardo {\it et al.},
  arXiv:1106.3014 [astro-ph.CO];
  J.~I.~Collar,
  arXiv:1106.3559 [astro-ph.CO];
  SIMPLE~Collaboration,
  arXiv:1107.1515 [astro-ph.CO].

\end{thebibliography}



\end{document}